\documentclass[aps,prl,twocolumn,superscriptaddress]{revtex4-1}
\usepackage{hyperref}
\usepackage{graphicx}  
\usepackage{dcolumn}   
\usepackage{bm}       
\usepackage{amssymb,amsmath}   
\usepackage[usenames,dvipsnames]{color}
\hypersetup{
  colorlinks = true,
  citecolor = {red},
  urlcolor = {blue}
}

\newcommand{\avg}[1]{\langle #1 \rangle}

\makeatletter
\renewcommand*\env@matrix[1][\arraystretch]{
  \edef\arraystretch{#1}
  \hskip -\arraycolsep
  \let\@ifnextchar\new@ifnextchar
\array{*\c@MaxMatrixCols c}}
\makeatother

\begin{document}
\title{Dynamic Riemannian Geometry of the Ising Model}
\author{Grant M. Rotskoff}
\email{rotskoff@berkeley.edu} 
\affiliation{Biophysics Graduate Group, University of California, Berkeley, CA 94720, USA}
\author{Gavin E. Crooks}
\affiliation{Molecular Biophysics Division, Lawrence Berkeley National Laboratory, Berkeley, CA 94720, USA}
\affiliation{Kavli Energy NanoSciences Institute, Berkeley, CA 94720, USA}

\begin{abstract}
A general understanding of optimal control in non-equilibrium systems would illuminate the operational principles of biological and artificial nanoscale machines.
Recent work has shown that a system driven out of equilibrium by a linear response protocol is endowed with a Riemannian metric related to generalized susceptibilities, and that geodesics on this manifold are the non-equilibrium control protocols with the lowest achievable dissipation.
While this elegant mathematical framework has inspired numerous studies of exactly solvable systems, no description of the thermodynamic geometry yet exists when the metric cannot be derived analytically. 
Herein, we numerically construct the dynamic metric of the 2D Ising model in order to study optimal protocols for reversing the net magnetization.
\end{abstract}

\maketitle

\noindent \emph{Introduction.---} 
At the nanoscale biological systems operate in a heterogeneous, fluctuating environment.
Nevertheless, life has been overwhelmingly successful at constructing machines that are fast~\cite{Lan2012}, accurate~\cite{Murugan2012,Murugan2014}, and efficient~\cite{Yasuda1998}. 
The recent development of techniques for nanoscale manipulation and design~\cite{Blickle2012, Collin2005, Toyabe2010}, alongside theoretical advances in non-equilibrium statistical mechanics~\cite{Spinney2013}, has given us a new set of tools with which to probe the thermodynamics of small systems operating out of equilibrium.
With these tools we can uncover the principles that have guided the evolution of molecular machines and shed light on the design of optimal nanoscale devices.

\begin{figure}[t!]
\includegraphics[width=\linewidth]{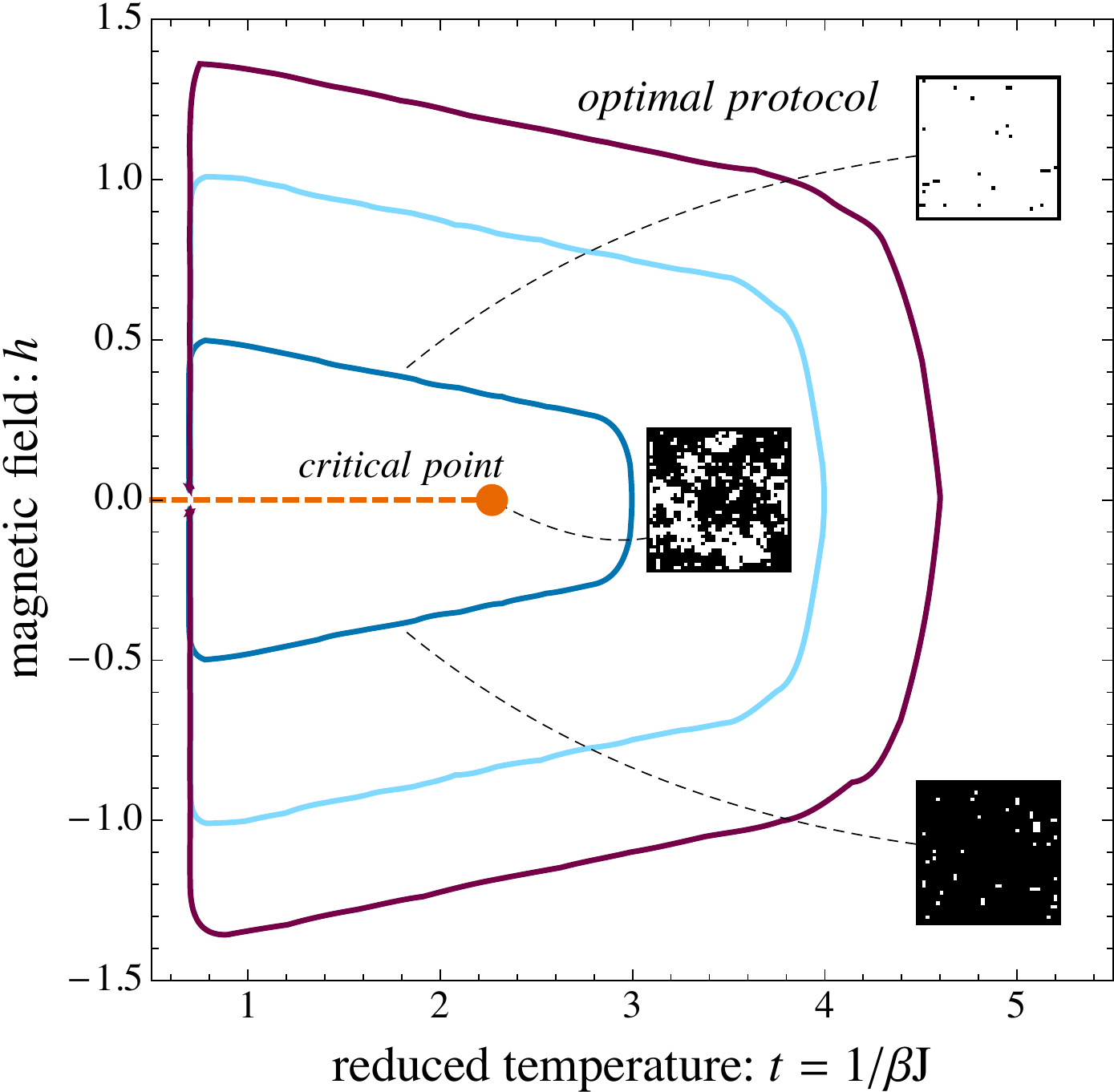}
\caption{Minimum dissipation, finite-time protocols for reversing the magnetization of the 2D Ising model with initial and final conditions below the critical temperature, $t_{\rm C}\approx 2.269.$ The outermost protocol is unconstrained, whereas the inner two protocols have a constraint on the maximum temperature. We control the external field $h$ and the spin-spin coupling constant $J$ as a function of time. Initially, the protocols ramp up the external field followed by a temperature increase as the field is turned off. Low dissipation protocols circumscribe the critical region to avoid large spatial and temporal correlations near the second order phase transition. The first order phase transition ($h=0,\ t<t_{\rm C}$) is shown as a dashed line ending at the critical point.}
\label{fig:geodesic}
\end{figure}

In the last several years, a geometric approach to non-equilibrium thermodynamics~\cite{Weinhold1975a, Salamon1980, Ruppeiner1979} has been extended to nanoscale systems~\cite{ Crooks2007c, Sivak2012b, Muratore-Ginanneschi2013}. 
To address the question of energy efficiency in stochastic machines, we imagine exercising control over a system by adjusting external parameters over some finite amount of time.
A typical control parameter might be the location of a harmonic potential trapping an optical bead or the magnitude of an applied magnetic field. 
An \emph{optimal protocol} is a prescription for changing the control parameters as a function of time that minimizes the average energy dissipated to the environment.
In the linear response regime, the space of control parameters is endowed with a Riemannian metric. 
On this manifold, distance minimizing geodesics are the minimum dissipation protocols. 

While studying model systems helps us glean the general principles of non-equilibrium control, theoretical analysis has thus far been restricted to single-body systems with exactly solvable dynamics~\cite{Zulkowski2012,Zulkowski2014,Schmiedl2007,Deffner2013,Muratore-Ginanneschi2014,Zulkowski2015} or in which the dynamics is not incorporated~\cite{Hunter1993, Shenfeld2009}.
For most systems of interest we cannot compute the metric exactly.
In this paper, we study the 2D Ising model and develop methods to predict optimal protocols from numerical computation of the metric tensor, adapting techniques originally developed in computational geometry~\cite{Sethian1996,Kimmel1998}.

The Ising model is a cornerstone of statistical mechanics that captures the essential physics of a diverse set of systems including ferromagnets, liquid-vapor phase transitions, and lipid membranes~\cite{Chandler1987,Honerkamp2008}. 
By studying the Ising model, we gain insight into the unexplored consequences of non-linear dynamics and the presence of a phase transition on optimal protocols.
Control can be exercised by applying an external field, but also by varying the spin-spin coupling, as in heat assisted magnetic recording~\cite{Kryder2008}.
Applications include magnetic information storage technologies that rely on inverting the net magnetization of microscopic spin domains as well as technologies for ultra low energy computation, such as hybrid spintronics~\cite{Roy2011}. 
Low dissipation control of seemingly simple, stochastic systems, such as spins on a magnetic hard drive, has implications for the efficiency of nanodevices already in wide use.

\noindent \emph{Ising model.---} 
We will consider the problem of inverting the net magnetization of the 2D Ising model using the spin-spin coupling and external field as control parameters.
In doing so, we also demonstrate a general strategy for numerically computing optimal protocols. 
The system is governed by the standard Ising hamiltonian,
\begin{equation}\label{eq:ising}
H \bigl(\{s_i\},\lambda(t) \bigr) = h(t)\sum_{i=1}^n s_i + J(t) \sum_{\langle i,j \rangle} s_i s_j\ ,
\end{equation}
where $\avg{i,j}$ denotes a sum over all nearest neighbor pairs on the lattice, and the control parameter
\begin{math}
\lambda(t) = \bigl(\beta h(t), \beta J(t)\bigr),
\end{math}
varies the coupling,~$J$, and the external magnetic field,~$h$, with time.
Controlling the strength of the spin-spin coupling can also be implemented by varying the temperature.

\begin{figure} 
\includegraphics[width=\linewidth]{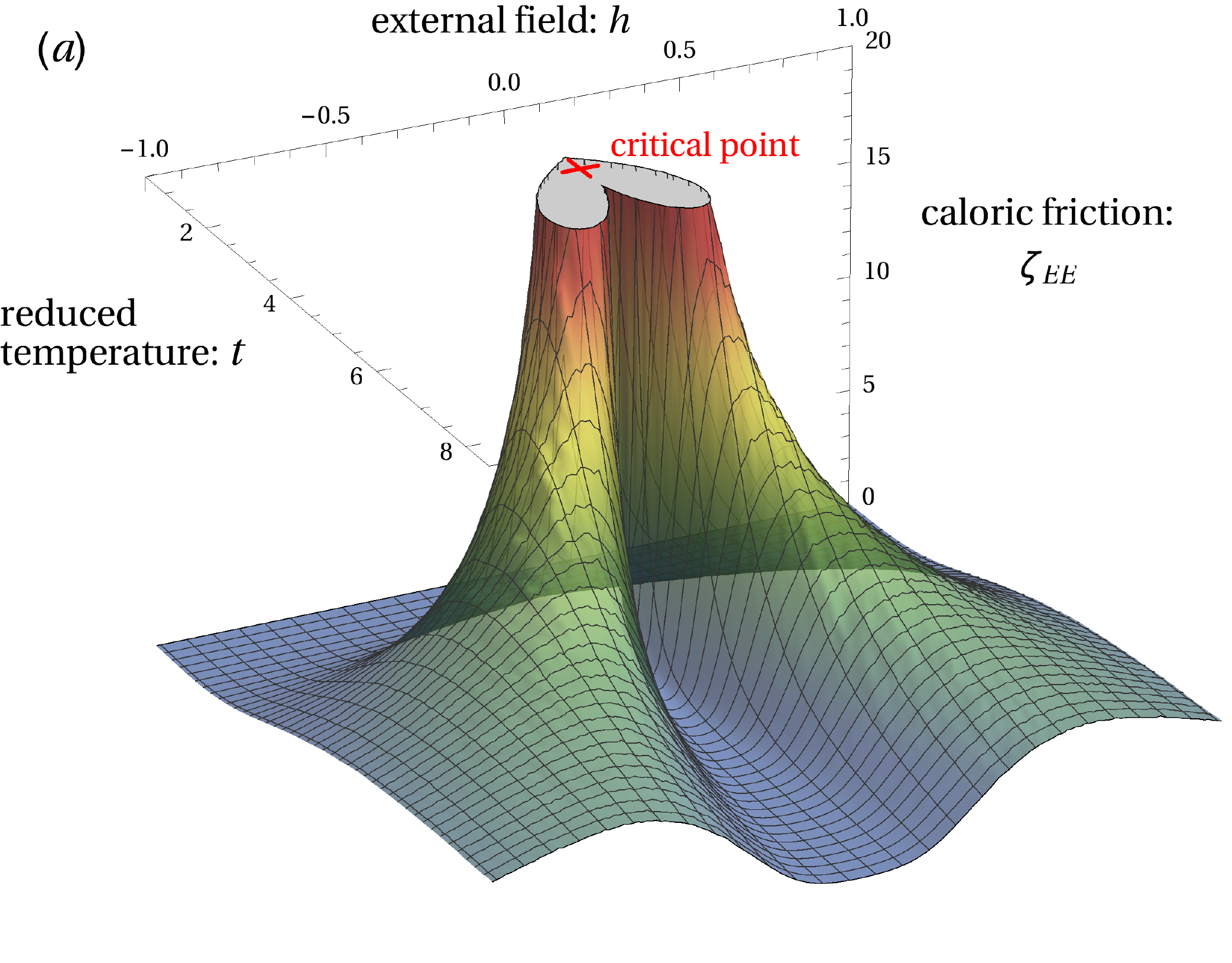}
\includegraphics[width=\linewidth]{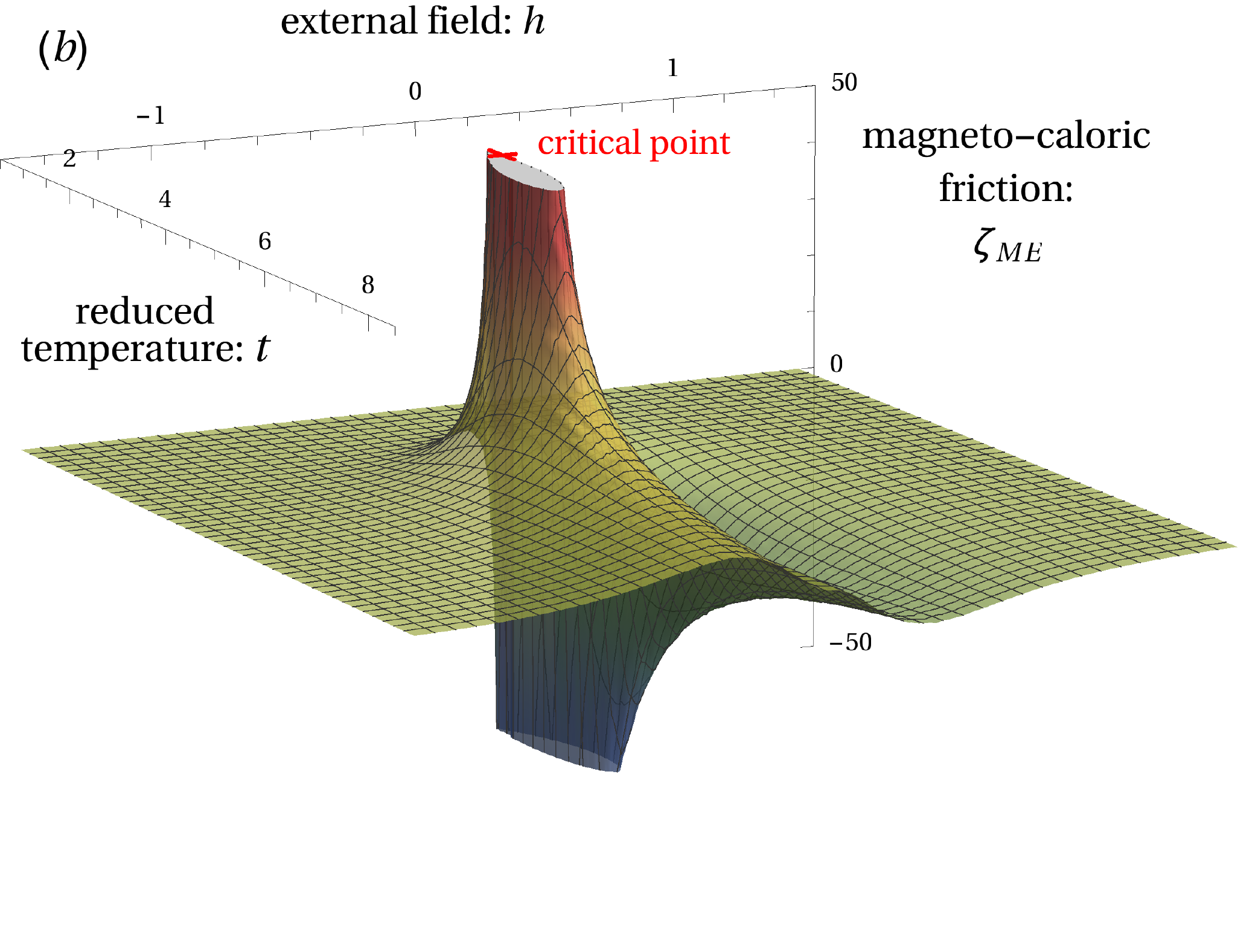}
\includegraphics[width=\linewidth]{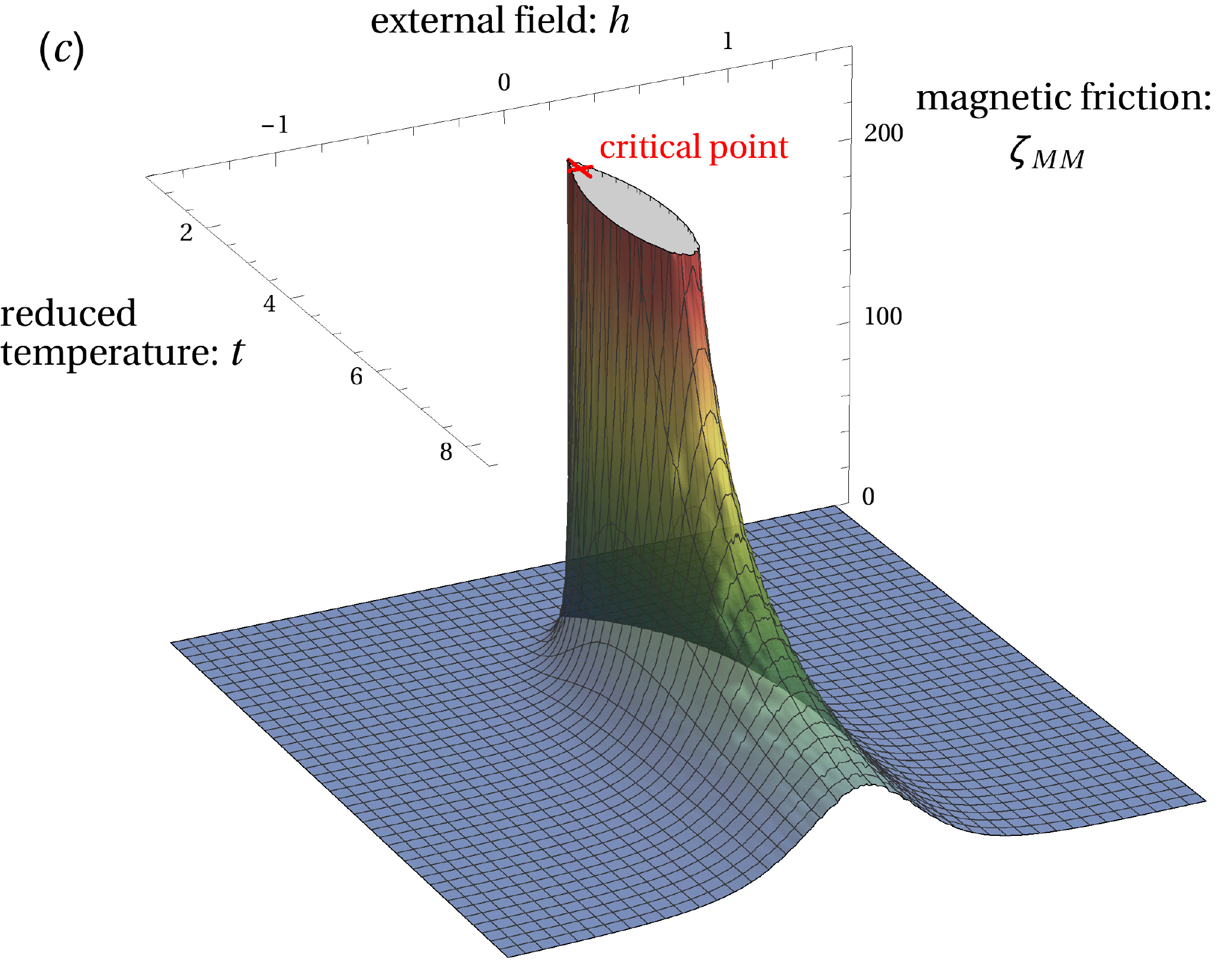}
\caption{The caloric $(a),$ magneto-caloric $(b),$ and magnetic $(c)$ friction coefficients of the 2D Ising model, as defined by Eq.~\eqref{eq:metric}, plotted in the magnetic field ($h$), temperature $T = 1/\beta J$ plane. 
Both relaxation times and static correlations diverge at the critical point which gives rise to the cusp in each of these plots. 
The friction coefficients are the matrix elements of a Riemannian metric with the property that geodesics minimize the average excess work that a protocol exercises over the system.}
\label{fig:metric}
\end{figure}

If we drive at a finite rate the system resists the changes in the control parameters.
In the linear regime, the friction,~$\zeta$~\cite{Sivak2012b}, that the protocol encounters is,
\begin{equation}\label{eq:metric}
\zeta_{ij}(\lambda(t)) = 
  \beta \int_0^\infty d\tau\ \bigl\langle \delta X_i(0) \delta X_j(\tau) \bigr\rangle_{\lambda(t)},
\end{equation}
where~$X_i$ is the conjugate force to the control parameter~$\lambda_i$ and $\delta X = X-\avg{X}.$
When we control the field and coupling, the conjugate forces are the net magnetization $M$ and internal energy $E$,
\begin{align}\label{eq:conjugate}
X_{\beta h}(t) &= \sum_{i=1}^n s_i \equiv M, \\
X_{\beta J}(t) &= \sum_{\langle i,j \rangle} s_i s_j \equiv E.
\end{align}
Similar expressions for the friction \eqref{eq:metric} arise in Kirkwood's linear response formula~\cite{Kirkwood1946,Sivak2012b} and also in the study of effective diffusion constants under coarse-graining~\cite{Berezhkovskii2011}.

\noindent{\emph{Thermodynamic geometry.---}}
The friction matrix \eqref{eq:metric} is a semi-Riemannian metric tensor---it is a symmetric, positive semi-definite, bilinear form. 
This metric defines the distance along a protocol~$\lambda$,
\begin{equation}
\mathcal{L}[\lambda(t)] 
      = \int_{\lambda} \dot{\lambda}^i \zeta_{ij} \dot{\lambda}^j,
\end{equation}
and the distance along an optimal protocol sets a lower bound on the excess work exercised by the controller over the system~\cite{Salamon1983,Sivak2012b},
\begin{equation}
\label{eq:workbound}
\Delta t \avg{W_{\rm ex}} \geq \mathcal{L}^2.
\end{equation}

For any protocol, equality between the divergence $\Delta t \avg{W_{\rm ex}}$ and the squared thermodynamic length $\mathcal{L}^2$ is achieved when the excess power is constant over the duration of the protocol.
As a result, the path of an optimal protocol does not depend its duration~\cite{Salamon1983,Crooks2007c}. 

Exact equations for the relaxation of $M$ and $E$ are not known in general, so we must approximate the metric using simulations. 
We discretize the parameter space and at each point we compute the time correlation matrix for the conjugate forces, 
\begin{equation}\label{eq:covariance}
  \begin{pmatrix}[1.5]
\bigl\langle \delta{X}_{\beta h}(0)\ \delta{X}_{\beta h}(\tau) \bigr\rangle & 
\bigl\langle \delta{X}_{\beta h}(0)\ \delta{X}_{\beta J}(\tau) \bigr\rangle \\
\bigl\langle \delta{X}_{\beta J}(0)\ \delta{X}_{\beta h}(\tau) \bigr\rangle & 
\bigl\langle \delta{X}_{\beta J}(0)\ \delta{X}_{\beta J}(\tau) \bigr\rangle 
\end{pmatrix}.
\end{equation}
The time correlation functions are estimated with Markov Chain Monte Carlo simulations on a $128$ by $128$ square lattice of Ising spins with Glauber dynamics \cite{Glauber1963}. 
We compared our results to a $256$ by $256$ system to ensure there were no significant finite size effects, aside from finite size scaling.
Integrating the time correlation function~\eqref{eq:covariance} to infinite time yields the friction coefficient $\zeta_{ij}$ \eqref{eq:metric} at each point in the parameter space. 
In practice, correlations decay exponentially and the friction tensor can be accurately estimated, except very near the critical point. 

Once the metric is known on a subspace of the parameter manifold, we recast the problem of approximating geodesic distances in terms of an \emph{eikonal} equation, 
\begin{math}
| \nabla T(t,h) | = 1 / F(t,h),
\end{math}
a partial differential equation commonly used to study wave propagation~\cite{Sethian1996}. The field~$F$ is the instantaneous speed of a wavefront and~$T$ represents the arrival time of the wave. In our case,~$F$ is the linearized Riemannian distance between neighboring points $\lambda_0$ and $\lambda_1,$ 
\begin{equation}\label{eq:linearized}
\nonumber
d(\lambda_0,\lambda_1) = 
	\sqrt{\frac{1}{2}(\lambda_1-\lambda_0)^{T}  \bigl(\zeta(\lambda_0) + \zeta(\lambda_1)\bigr) (\lambda_1-\lambda_0)}.
\end{equation}
We expect linearization to be robust so long as the discretization is sufficiently fine. In the vicinity of the critical point, we computed the friction tensor on a finer mesh. 

We used the ``fast marching method''~\cite{Sethian1996} to find numerical solutions to the \emph{eikonal} equation. 
This algorithm approximates continuous geodesic paths, as shown in Fig.~\ref{fig:geodesic}, given discrete knowledge of the distance between neighboring points ~\cite{Kimmel1998}.
A geodesic path in the parameter space travels backwards along the gradient of~$T.$ 
After computing the arrival time field~$T$ for geodesics initiated from some initial point $\lambda_0$, we can solve a first order differential equation, to find a geodesic path between $\lambda_0$ and $\lambda_1.$ 
Given the metric, we can rapidly calculate optimal protocols between any two points.

\noindent \emph{Ising metric.---}
In Fig.~\ref{fig:metric} we plot each of the components of the friction tensor.
The caloric friction coefficient $\zeta_{EE}$ is the time autocorrelation of the internal energy. 
At each point in parameter space, this friction can be written as $\zeta_{EE} = \tau_{EE}\avg{(\delta E)^2} = \tau_{EE}k_{\rm B} t^2 C,$ the product of the heat capacity $C$ and an effective timescale for the relaxation of the energy. 
Similarly, the cross-correlation of the magnetization and internal energy, the magneto-caloric friction $\zeta_{ME} = \zeta_{EM} = \tau_{EM}k_{\rm B}t M_{t}$ is proportional to the magneto-caloric coefficient $M_t$, and the autocorrelation of the magnetization, the magnetic friction $\zeta_{MM} = \tau_{MM}\chi$ is proportional to the magnetic susceptibility $\chi.$

Both static correlations and relaxation timescales diverge near the critical point of the Ising model.
These two effects compound to produce a singularity of the metric where all three components of friction tensor also diverge.
The friction coefficients decay according to characteristic power laws in neighborhoods surrounding the critical point \cite{Cardy1996}. 
Correlations are also large exactly at the first order phase transition along the line $h=0,$ $t<t_{\rm C}.$ 
However, spontaneous magnetization reversal is rarely observed in simulations under single spin flip dynamics.
Below the critical temperature $t_{\rm C}$ with $h\neq 0$, relaxation times are fast and fluctuations are negligible, which results in small values for each component of the friction tensor.

The geometry of the supercritical region is more intricate. 
The caloric friction, Fig.~\ref{fig:metric}~(a), exhibits symmetric ridges that correspond to maxima in the heat capacity and are reminiscent of ``Widom lines'' in supercritical fluids~\cite{Xu2005}.
Along these ridges we observe large, slowly relaxing spin domains.
The magneto-caloric friction, Fig.~\ref{fig:metric}~(b), is antisymmetric about $h=0$ due to the antisymmetry in the net magnetization.
The magnetic friction, Fig.~\ref{fig:metric}~(c), is large for an extended region above the critical temperature. 
At very high temperatures, all the components of the metric are again small due to neglible spin-spin couplings.

\begin{figure}[t!]
\includegraphics[width=\linewidth]{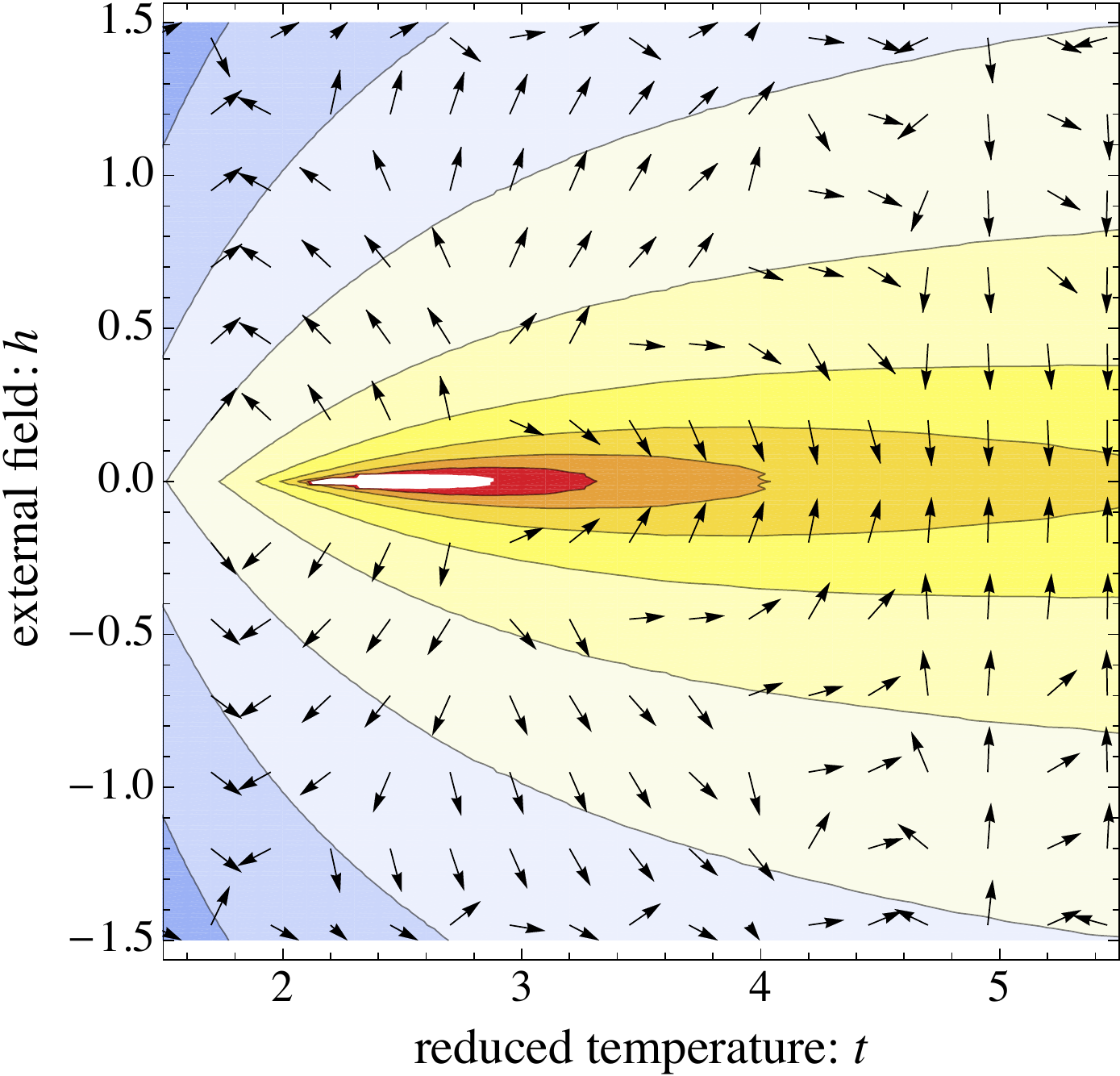}
\caption{
  Black arrows show tangents vectors of geodesic paths passing through $h=0, t=4.6,$ where the optimal protocol plotted in Fig.~\ref{fig:geodesic} crosses the supercritical line.
  Optimal protocols follow these geodesic flows. 
  Starting below the critical temperature, geodesics flow towards high field and low temperatures before raising temperature and subsequently reducing the field.
  Contours show $\log {\rm Tr}\ \zeta.$
  }
\label{fig:isolines}
\end{figure}

\noindent \emph{Optimal protocols.---}
Protocols, as shown in Fig.~\ref{fig:isolines}, clearly avoid the critical point by curving around this feature of the phase diagram due to the high friction in this region.
Passing directly through the first order phase transition, even in a finite time, also has a high dissipation cost.
Overcoming the broken symmetry requires nucleation of a domain of opposite spin, which can then grow to reverse the net magnetization.
Nucleation can be accelerated by applying a large field, but this results in a proportionally higher dissipation when the spins reverse.

At low temperatures, excitations are small and local, which leads to low friction (See configurations in Fig.~\ref{fig:geodesic}).
As a result, the protocols are weakly constrained below the critical temperature.
Similarly, in the high temperature limit, the vanishing spatial and temporal correlations result in low friction and weakly constrained protocols.
Only at intermediate temperatures does higher friction imposes tight constraints on the minimum dissipation paths.

Optimal protocol for reversing the magnetization are plotted in~Fig.~\ref{fig:geodesic}.
Counterintuitively, the magnetic field is first applied in the direction of the spontaneous magnetization.
Because the friction coefficients are small in the low temperature region, aligning the spins at the outset minimizes the overall contribution to the dissipation by dampening fluctuations as the temperature of the system is brought above the critical temperature.
The direction of the field is then reversed, but since the value of the magnetic friction coefficient is large along the zero external field line, as shown in Fig.~\ref{fig:metric}~(c), crossing between positive and negative field must be performed slowly. 
The protocol is symmetric about zero field due to the underlying symmetry of the model, thus we reduce the temperature and finally turn off the field. 

\noindent \emph{Discussion.---} 
Optimal protocols depend on what we can control. 
For instance, given spatial control of the external field, the minimum dissipation protocol may involve flipping spins at the boundary of a domain.
High dimensional parameter spaces will require different approaches to calculating geodesics.
Analogous problems in transition state theory have been addressed using the string method~\cite{E2002} and path sampling~\cite{Bolhuis2002}.

Non-equilibrium nanoscale machines need to be designed for objectives beyond low dissipation.
If speed is the objective, the bound in Eq.~\eqref{eq:workbound} can be used to minimize the total duration of the protocol, while keeping the average dissipation fixed.
Supercritical heat engines~\cite{Dobashi1998} and magnetic refrigerators~\cite{Manosa2013} could also be studied using the Ising model, but in these cases the objective is to efficiently transfer energy around a thermodynamic cycle.
In such cases, we will have to include additional constraints when seeking efficient control.
There may also be practical limits on the range of the control parameters. 
As an illustration, Fig.~\ref{fig:geodesic} shows minimum dissipation protocols where the maximum temperature is constrained.
The optimal protocols we have predicted are weakly constrained where the manifold is flat, affording tremendous flexibility to the controller. 
Where the metric changes rapidly, protocols are tightly constrained and external control must be precise.

\subsection{Acknowledgments}
\begin{acknowledgments}
G.M.R. acknowledges enlightening discussions with James Sethian concerning the use and implementation of the fast marching method. They would also like to thank David Sivak for useful preliminary conversations. G.M.R. wishes to thank support from the NSF graduate research fellowship. This work was supported in part by the US Army Research Laboratory and the US Army Research Office under Contract No.~W911NF-13-1-0390. 
\end{acknowledgments}
\bibliography{refs,GECLibrary}
\end{document}